\documentclass{article}
\usepackage{geometry}
\usepackage{graphicx}
\usepackage{amssymb, amsmath}
\usepackage{subfigure}
\usepackage{color}
\usepackage{hyperref}
\usepackage[accepted]{icml2015}
\usepackage{tikz}

\usetikzlibrary{intersections, backgrounds}

\definecolor{light}{RGB}{199, 153, 199}
\definecolor{dark}{RGB}{180, 31, 180}
\definecolor{gray80}{gray}{0.8}

\graphicspath{{figures/}}

\begin{document}

\icmltitlerunning{The Fundamental Incompatibility of Hamiltonian Monte Carlo and Data Subsampling}

\twocolumn[
\icmltitle{The Fundamental Incompatibility of \\ Hamiltonian Monte Carlo and Data Subsampling}

\icmlauthor{Michael Betancourt}{betanalpha@gmail.com}
\icmladdress{Department of Statistics, University of Warwick, Coventry, UK CV4 7AL}

\icmlkeywords{Hamiltonian Monte Carlo, Symplectic Integrators, Subsampled Data}

\vskip 0.3in
]

\begin{abstract}
Leveraging the coherent exploration of Hamiltonian flow, Hamiltonian Monte Carlo 
produces computationally efficient Monte Carlo estimators, even with respect to
complex and high-dimensional target distributions.  When confronted with data-intensive 
applications, however, the algorithm may be too expensive to implement, leaving us to 
consider the utility of approximations such as data subsampling.  In this paper I demonstrate
how data subsampling fundamentally compromises the efficient exploration of Hamiltonian 
flow and hence the scalable performance of Hamiltonian Monte Carlo itself.
\end{abstract}

With the preponderance of applications featuring enormous data sets,
methods of inference requiring only subsamples of data are becoming 
more and more appealing.  Subsampled Markov Chain Monte Carlo 
algorithms, \cite{NeiswangerEtAl:2013, WellingEtAl:2011}, are particularly
desired for their potential applicability to most statistical models.  Unfortunately, 
careful analysis of these algorithms reveals unavoidable biases unless 
the data are \textit{tall}, or highly redundant 
\cite{BardenetEtAl:2014, TehEtAl:2014, VollmerEtAl:2015}.  
The utility of these subsampled algorithms is then a consequence 
of not only the desired accuracy and also the particular model and data
under consideration, which can severely restrict practicality.

Recently \cite{ChenEtAl:2014} considered subsampling within Hamiltonian
Monte Carlo \cite{DuaneEtAl:1987, Neal:2011, BetancourtEtAl:2014} and 
demonstrated that the biases induced by naive subsampling lead to 
unacceptably large biases.  Ultimately the authors rectified this bias by
sacrificing the coherent exploration of Hamiltonian flow for a diffusive correction, 
fundamentally compromising the scalability of the algorithm with respect to
the complexity of the target distribution.  An algorithm scalable with respect to
both the size of the data and the complexity of the target distribution would
have to maintain the coherent exploration of Hamiltonian flow while 
subsampling and, unfortunately, these objectives are mutually exclusive in
general.

In this paper I review the elements of Hamiltonian Monte Carlo critical to its
robust and scalable performance in practice and demonstrate how different 
subsampling strategies all compromise those properties and consequently 
induce poor performance. 

\section{Hamiltonian Monte Carlo in Theory}

Hamiltonian Monte Carlo utilizes deterministic, measure-preserving maps to 
generate efficient Markov transitions.  Formally, we begin by complementing 
a target distribution,
\begin{equation*}
\varpi \propto \exp \! \left[ - V ( q ) \right] \mathrm{d}^{n} q,
\end{equation*}
with a conditional distribution over auxiliary \textit{momenta} parameters,
\begin{equation*}
\varpi_{q} \propto \exp \! \left[ - T (p, q) \right] \mathrm{d}^{n} p.
\end{equation*}
Together these define a joint distribution,
\begin{align*}
\varpi_{H} 
&\propto \exp \! \left[ - \left( T (q, p) + V (q) \right) \right] \mathrm{d}^{n} q \, \mathrm{d}^{n} p
\\
&\propto \exp \! \left[ - H (q, p) \right] \mathrm{d}^{n} q \, \mathrm{d}^{n} p,
\end{align*}
and a \textit{Hamiltonian system} corresponding to the \textit{Hamiltonian},
$ H (q, p) $.  We refer to $T (q, p)$ and $V (q)$ as the \textit{kinetic energy} 
and \textit{potential energy}, respectively.

The Hamiltonian immediately defines a \textit{Hamiltonian flow} on the joint
space,
\begin{align*}
\phi^{H}_{t} : (q, p) &\rightarrow (q, p), \forall t \in \mathbb{R}
\\
\phi^{H}_{t} \circ \phi^{H}_{s} &= \phi^{H}_{s + t},
\end{align*}
which exactly preserves the joint distribution,
\begin{equation*}
\left( \phi^{H}_{t} \right)_{*} \varpi_{H} = \varpi_{H}.
\end{equation*}
Consequently, we can compose a Markov chain by sampling the auxiliary momenta, 
\begin{equation*}
q \rightarrow (q, p), \, p \sim \varpi_{q},
\end{equation*}
applying the Hamiltonian flow,
\begin{equation*}
(q, p) \rightarrow \phi^{H}_{t} (q, p)
\end{equation*}
and then projecting back down to the target space,
\begin{equation*}
(q, p) \rightarrow q.
\end{equation*}
By construction, the trajectories generated by the Hamiltonian flow 
explore the level sets of the Hamiltonian function, which shadow the
probability mass of the joint distribution.  Because these level sets
can also span large volumes of the joint space, sufficiently-long 
trajectories can yield transitions far away from the initial state of the 
Markov chain, drastically reducing autocorrelations and producing 
computationally efficient Monte Carlo estimators.

When the kinetic energy does not depend on position we say that
the Hamiltonian is \textit{separable}, $H (q, p) = T (p) + V (q)$,  and
the exact Hamiltonian flow can be generated by the \textit{Hamiltonian operator}, 
$\hat{H}$,
\begin{equation*}
\phi^{H}_{\tau} = e^{\tau \hat{H}},
\end{equation*}
where
\begin{align*}
\hat{H} 
&= 
\frac{ \partial H }{ \partial p } \frac{ \partial }{ \partial q}
- \frac{ \partial H }{ \partial q } \frac{ \partial }{ \partial p}
\\
&= 
\frac{ \partial T }{ \partial p } \frac{ \partial }{ \partial q}
- \frac{ \partial V }{ \partial q } \frac{ \partial }{ \partial p}
\\
&\equiv
\quad\, \hat{T} \quad + \quad \hat{V} \quad.
\end{align*}
In this paper I consider only separable Hamiltonians, although
the conclusions also carry over to the non-seperable Hamiltonians,
for example those arising in
Riemannian Hamiltonian Monte Carlo \cite{GirolamiEtAl:2011}.

\section{Hamiltonian Monte Carlo in Practice}

The biggest challenge of implementing Hamiltonian Monte Carlo is that the
Hamiltonian operator is rarely calculable in practice and we must instead resort 
to approximate integration of the Hamiltonian flow.  \textit{Symplectic integrators}, 
which yield numerical trajectories that closely track the true trajectories, are  
of particular importance to any high-performance implementation. 

An especially transparent strategy for constructing symplectic integrators is to split 
the Hamiltonian into terms with soluble flows which can then be composed together
\cite{LeimkuhlerEtAl:2004, HairerEtAl:2006}.  For example, consider the symmetric 
\textit{Strang} splitting,
\begin{equation*}
\phi^{V}_{\frac{\epsilon}{2}} \circ 
\phi^{T}_{\epsilon} \circ 
\phi^{V}_{\frac{\epsilon}{2}}
=
e^{\frac{\epsilon}{2} \hat{V} } \circ 
e^{\epsilon \hat{T} } \circ 
e^{\frac{\epsilon}{2} \hat{V} },
\end{equation*}
where $\epsilon$ is a small interval of time known as the \textit{step size}.
Appealing to the Baker-Campbell-Hausdorff formula, this symmetric composition yields
\begin{align*}
\phi^{V}_{\frac{\epsilon}{2}} \circ 
\phi^{T}_{\epsilon} \circ 
\phi^{V}_{\frac{\epsilon}{2}}
\\
& \hspace{-12mm} =
e^{\frac{\epsilon}{2} \hat{V} } \circ 
e^{\epsilon \hat{T} } \circ 
e^{\frac{\epsilon}{2} \hat{V} }
\\
& \hspace{-12mm} =
e^{\frac{\epsilon}{2} \hat{V} } \circ 
\exp \left( \epsilon \hat{T} + \frac{\epsilon}{2} \hat{V} 
+ \frac{\epsilon^{2}}{4} \left[ \hat{T}, \hat{V} \right] \right)
+ \mathcal{O} \! \left( \epsilon^{3} \right)
\\
& \hspace{-12mm} =
\exp \left( 
\frac{\epsilon}{2} \hat{V} + \epsilon \hat{T} + \frac{\epsilon}{2} \hat{V} 
+ \frac{\epsilon^{2}}{4} \left[ \hat{T}, \hat{V} \right]
\right.
\\
& \hspace{0mm} \left.
+ \frac{1}{2} \left[ \frac{\epsilon}{2} \hat{V}, \epsilon \hat{T} + \frac{\epsilon}{2} \hat{V} 
+ \frac{\epsilon^{2}}{4} \left[ \hat{T}, \hat{V} \right] \right]
\right)
+ \mathcal{O} \! \left( \epsilon^{3} \right)
\\
\\
& \hspace{-12mm} =
e^{ \epsilon \hat{H} }
+ \mathcal{O} \! \left( \epsilon^{3} \right).
\end{align*}
Composing this symmetric composition with itself $L = \tau / \epsilon$ times results in 
a symplectic integrator accurate to second-order in the step size for any finite integration 
time, $\tau$,
\begin{align*}
\phi^{\widetilde{H}}_{\epsilon, \tau}
&\equiv
\left( \phi^{V}_{\frac{\epsilon}{2}} \circ 
\phi^{T}_{\epsilon} \circ 
\phi^{V}_{\frac{\epsilon}{2}} \right)^{L}
\\
&=
\left( e^{ \epsilon \hat{H} }
+ \mathcal{O} \! \left( \epsilon^{3} \right) \right)^{L}
\\
&=
e^{ \left( L \epsilon \right) \hat{H} }
+ \left( L \epsilon \right) \mathcal{O} \! \left( \epsilon^{2} \right)
\\
&=
e^{ \tau \hat{H} }
+ \tau \mathcal{O} \! \left( \epsilon^{2} \right)
\\
&=
e^{ \tau \hat{H} }
+ \mathcal{O} \! \left( \epsilon^{2} \right).
\end{align*}
Remarkably, the resulting numerical trajectories are confined to the level sets
of a \textit{modified Hamiltonian} given by an $\mathcal{O} \! \left( \epsilon^{2} \right)$
perturbation of the exact Hamiltonian \cite{HairerEtAl:2006, BetancourtEtAl:2014b}.

Although such symplectic integrators are highly accurate, they still introduce an
error into the trajectories that can bias the Markov chain and any resulting
Monte Carlo estimators.  In practice this error is typically compensated with
the application of a Metropolis acceptance procedure, accepting a point along
the numerical trajectory only with probability
\begin{equation*}
a (p, q) = \min \left(1, 
\exp \! \left( H (q, p) - H \circ \phi^{\widetilde{H}}_{\epsilon, \tau} (q, p) \right) \right).
\end{equation*}

A critical reason for the scalable performance of such an implementation of 
Hamiltonian Monte Carlo is that the error in a symplectic integrator scales 
with the step size, $\epsilon$.  Consequently a small bias or a large acceptance 
probability can be maintained by reducing the step size, regardless of the complexity
or dimension of the target distribution \cite{BetancourtEtAl:2014b}.  If the symplectic
integrator is compromised, however, then this scalability and generality is lost.

\section{Hamiltonian Monte Carlo With Subsampling}

A common criticism of Hamiltonian Monte Carlo is that in data-intensive applications 
the application of potential energy operator,
\begin{equation*}
\hat{V} = - \frac{ \partial V }{ \partial q } \frac{ \partial }{ \partial p},
\end{equation*}
and hence the simulation of numerical trajectories, can become infeasible given the 
expense of the gradient calculations.  This expense has fueled a variety of modifications 
of the algorithm aimed at reducing the cost of the potential energy operator, often by 
any means necessary.

An increasingly popular strategy targets Bayesian applications where the data are 
independently and identically distributed.  In this case the posterior can be manipulated
into a product of contributions from each subset of data, and the potential energy
likewise decomposes into a sum,
$V (q) = \sum_{j = 1}^{J} V_{j} (q)$,
where each $\left\{ V_{j} \right\}$ depends on only a single subset.  This decomposition 
suggests algorithms which consider not the entirety of the data and the full potential
energy, $V$, but rather only a few subsets at a time.

The performance of any such subsampling method depends critically on the details
of the implementation and the structure of the data itself.  Here I consider the performance 
of two immediate implementations, one based on subsampling the data in between 
Hamiltonian trajectories and one based on subsampling the data within a single trajectory.  
Unfortunately, the performance of both methods leaves much to be desired.

\subsection{Subsampling Data In Between Trajectories}

Given any subset of the data, we can approximate the potential energy as 
$V \approx J \, V_{j}$ and then generate trajectories corresponding to the flow of the 
approximate Hamiltonian, $H_{j} = T + J \, V_{j}$.  In order to avoid parsing the entirely
of the data, the Metropolis acceptance procedure can be neglected and the 
corresponding samples left biased.

\begin{figure*}
\centering
\subfigure[]{
\begin{tikzpicture}[scale=0.25, thick]
  
  \draw [light, thick] plot [smooth, tension=1.1] coordinates {(0,5) (10, 10) (21, 8) (25, 8)}
  node[right] { Stochastic };

  \draw [dark, thick] plot [smooth, tension=1.1] coordinates {(0,0) (10, 7) (20, 5) (24, 5)}
  node[right] { Exact };
  
  \draw[->, color=gray80, thick] (10, 10) -- +(3, 3);
  \draw[->, color=gray80, thick] (10, 10) -- +(3, -3);
  \draw[->, color=gray80, thick] (10, 10) -- +(1, 4);
  \draw[->, color=gray80, thick] (10, 10) -- +(-1, 5);
  \draw[->, color=gray80, thick] (10, 10) -- +(1, -2);
  \draw[->, color=gray80, thick] (10, 10) -- +(-1, -2);
  
  \draw[->, color=black, thick] (10, 10) -- +(3, 0.5);
  
\end{tikzpicture}
}
\subfigure[]{
\begin{tikzpicture}[scale=0.25, thick]

  \draw [light, thick] plot [smooth, tension=0.8] coordinates {(0,5) (2, 6.75) (4, 8) (6, 8.95) 
  (8, 9.6) (10, 10) (12, 10.7) (14, 13) (16, 17)}
  node[right] { Stochastic };

  \draw [dark, thick] plot [smooth, tension=1.1] coordinates {(0,0) (10, 7) (20, 5) (24, 5)}
  node[right] { Exact };
  
  \draw[->, color=gray80, thick] (10, 10) -- +(1, 4);
  \draw[->, color=gray80, thick] (10, 10) -- +(-1, 5);
  \draw[->, color=gray80, thick] (10, 10) -- +(1, -2);
  
  \draw[->, color=black, thick] (10, 10) -- +(2, 2);
  
\end{tikzpicture}
}
\caption{The bias induced by subsampling data in Hamiltonian Monte Carlo depends 
on how precisely the gradients of the subsampled potential energies integrate to the 
gradient of the true potential energy.  (a) When the subsampled gradient is close to the 
true gradient, the stochastic trajectory will follow the true trajectory and the bias will be small. 
(b) Conversely, if the subsampled gradient is not close to the true potential energy then the 
stochastic trajectory will drift away from the true trajectory and induce a bias.
Subsampling between trajectories requires that each subsampled gradient approximate
the true gradient, while subsampling within a single trajectory requires only that the
average of the subsampled gradients approximates the true gradient.  As the dimension of
the target distribution grows, however, an accurate approximation in either case
becomes increasingly more difficult unless the data become correspondingly more
redundant relative to the complexity of the target distribution.}
\label{fig:subsampled_gradients}
\end{figure*}
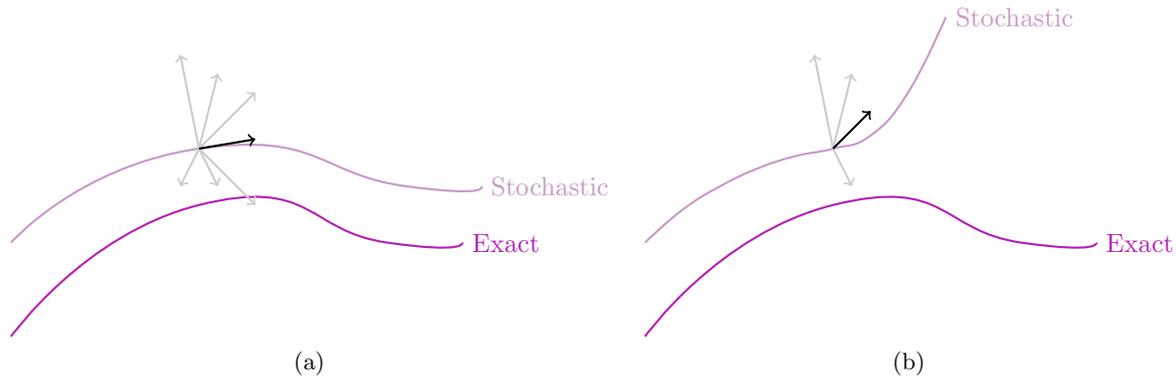

Unlike the numerical trajectories from the full Hamiltonian, these subsampled 
trajectories are biased away from the exact trajectories regardless of the chosen step size.  
In particular, the bias of each step,
\begin{align*}
e^{\frac{\epsilon}{2} I \hat{V}_{j} } \circ 
e^{\epsilon \hat{T} } \circ 
e^{\frac{\epsilon}{2} I \hat{V}_{j} }
&=
e^{ \epsilon \hat{H}_{j} }
+ \mathcal{O} \! \left( \epsilon^{3} \right)
\\
&=
e^{ \epsilon \hat{H} - \epsilon \widehat{ \Delta V}_{j} }
+ \mathcal{O} \! \left( \epsilon^{3} \right),
\end{align*}
where
\begin{equation} \label{eqn:between_bias}
\widehat{ \Delta V_{j} } = 
- \left( \frac{ \partial V }{ \partial q } - J \frac{ \partial V_{j} }{ \partial q } \right)
\frac{ \partial }{ \partial p },
\end{equation}
persists over an entire trajectory,
\begin{equation*}
\left( e^{\frac{\epsilon}{2} I \hat{V}_{j} } \circ 
e^{\epsilon \hat{T} } \circ 
e^{\frac{\epsilon}{2} I \hat{V}_{j} } \right)^{L}
=
e^{ \tau \left( \hat{H} - \widehat{ \Delta V}_{j} \right) }
+ \mathcal{O} \! \left( \epsilon^{2} \right).
\end{equation*}
As the dimension of the target distribution grows, the subsampled gradient,
$J \, \partial V_{j} / \partial q $, drifts away from the true gradient, 
$\partial V / \partial q $, unless the data become increasingly redundant.
Consequently the resulting trajectory introduces an irreducible bias into
the algorithm, similar in nature to the asymptotic bias seen in subsampled
Langevin Monte Carlo \cite{TehEtAl:2014, VollmerEtAl:2015}, which then 
induces either a vanishing Metropolis acceptance probability 
or highly-biased expectations if the Metropolis procedure is ignored outright 
(Figure \ref{fig:subsampled_gradients}).  

Unfortunately, the only way to decrease the dependency on redundant data
is to increase the size of each subsample, which immediately undermines
any computational benefits.

Consider, for example, a simple application where we target a one-dimensional
posterior distribution,
\begin{equation} \label{posterior}
p \! \left( \mu | \vec{x} \right) \propto p \! \left( \vec{x} | \mu \right) p \! \left( \mu \right),
\end{equation}
with the likelihood
\begin{equation*}
p \! \left( \vec{x} | \mu \right) = \prod_{n = 1}^{N} \mathcal{N} \! \left( x_{n} | \mu, \sigma^{2} \right)
\end{equation*}
and prior
\begin{equation*}
p \! \left( \mu \right) = \mathcal{N} \! \left( \mu | m, s^{2} \right). 
\end{equation*}
Separating the data into $J = N / B$ batches of size $B$ and decomposing the
prior into $J$ individual terms then gives
\begin{align*}
V_{j} &= \mathrm{const} + 
\frac{B}{N} \frac{ \sigma^{2} + N s^{2}  }{ \sigma^{2} s^{2} }
\\
& \quad
\times \left( \mu - 
\frac{ \left( \frac{1}{B} \sum_{n = (j - 1) B + 1}^{j B} x_{n} \right) N s^{2}  + m \sigma^{2} }
{ \sigma^{2} + N s^{2} } 
\right)^{2}.
\end{align*}
Here I take $\sigma = 2$, $m = 0$, $s = 1$, and generate $N = 500$ data points
assuming $\mu = 1$.

\begin{figure}
\centering
\subfigure[]{\includegraphics[width=1.75in]{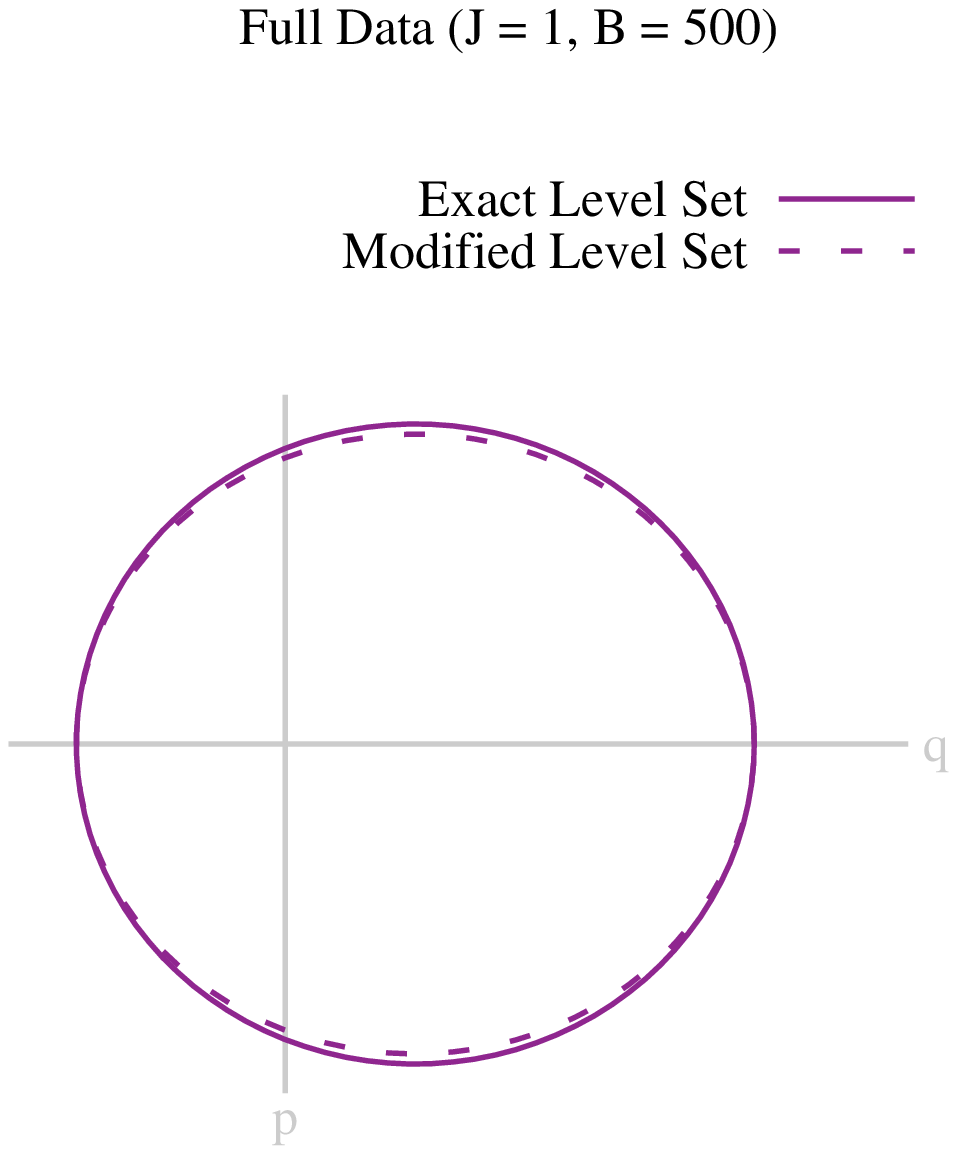}}
\subfigure[]{\includegraphics[width=1.75in]{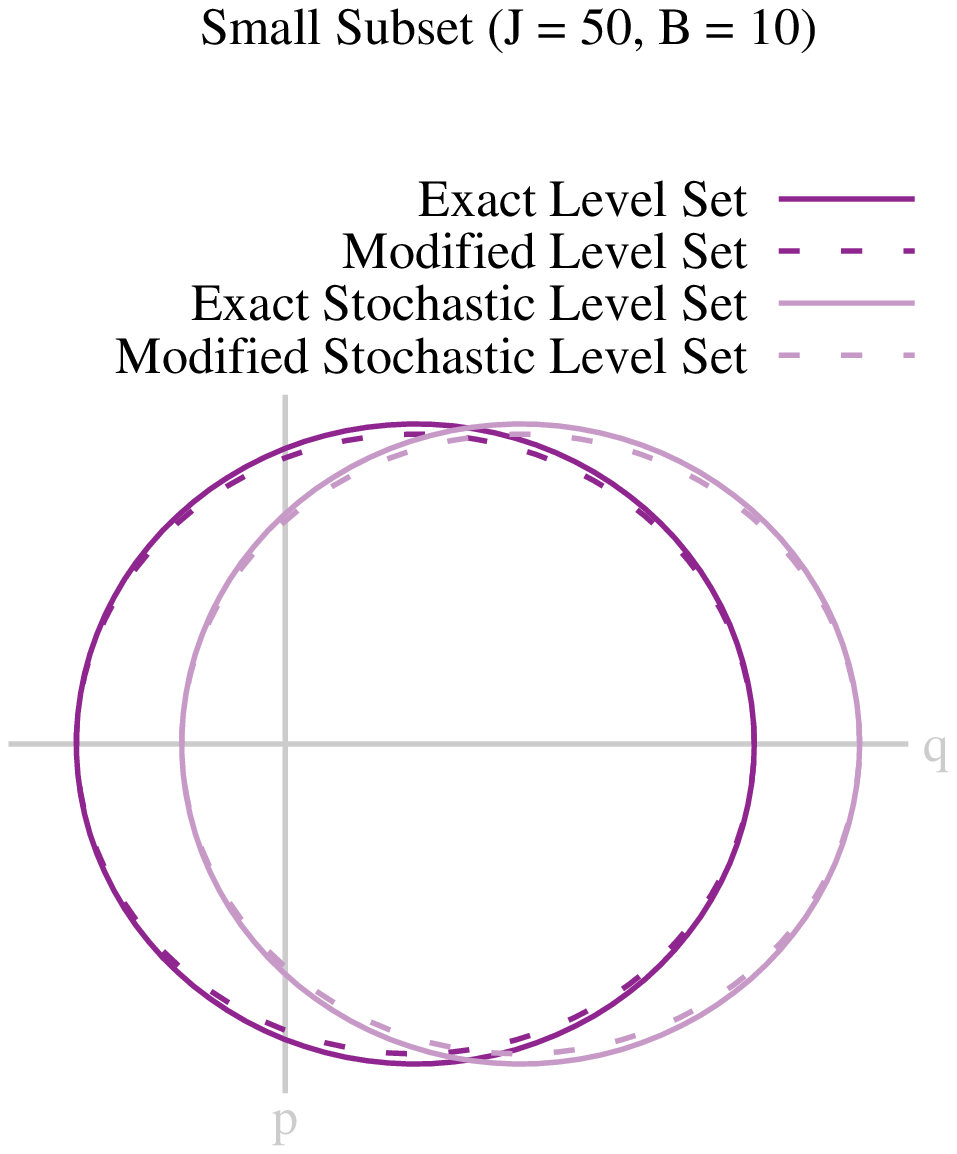}}
\subfigure[]{\includegraphics[width=1.75in]{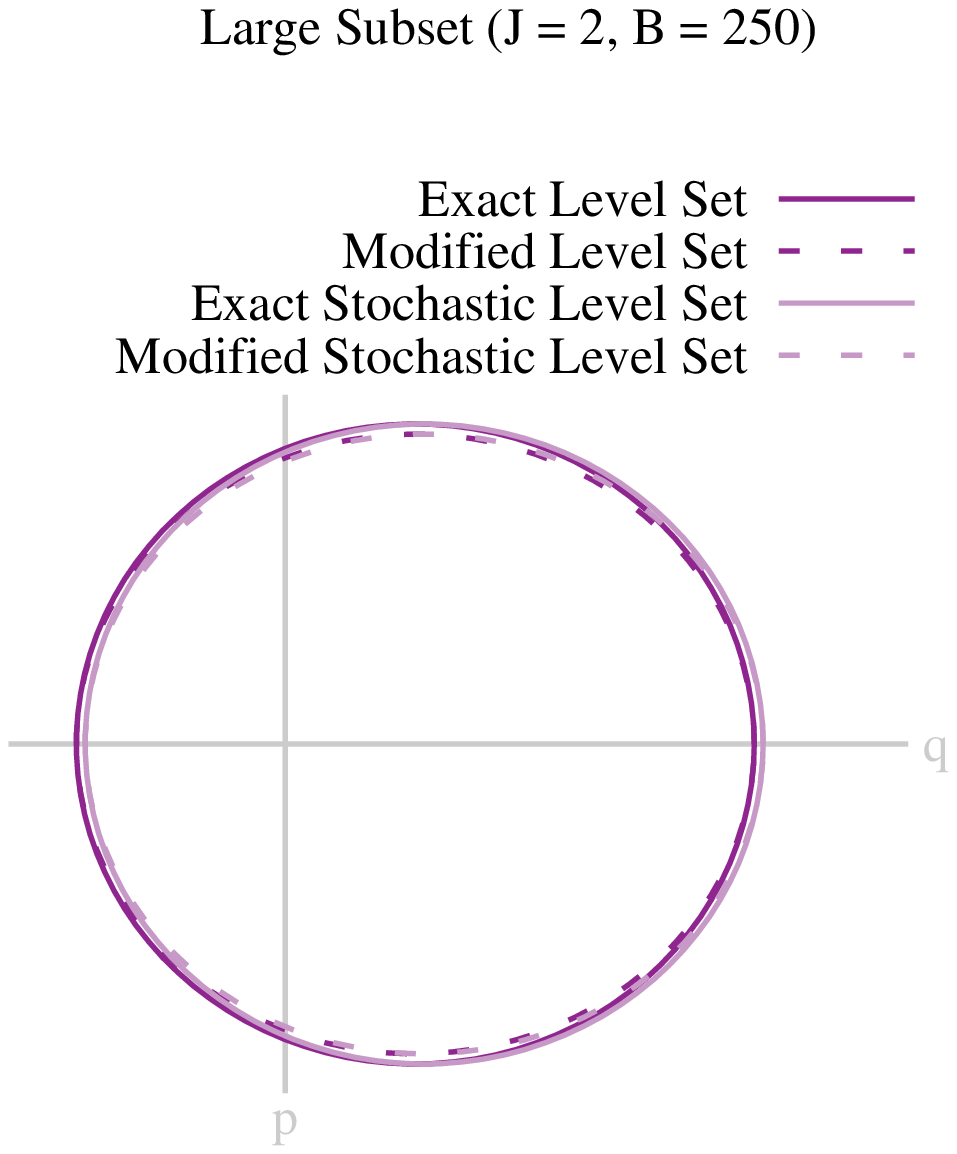}}
\caption{Even for the one-dimensional target distribution \eqref{posterior}, subsampling
data in between Hamiltonian trajectories introduces significant pathologies.
(a) When the full data are used, numerical Hamiltonian trajectories (dashed line) 
closely track the exact Hamiltonian trajectories (solid line).  Subsampling of the
data introduces a bias in both the exact trajectories and corresponding
numerical trajectories.  (b) If the size of each subsample is small then this bias 
is large, offsetting both the exact and numerical stochastic trajectories. 
(c) Only when the size of the subsamples approaches the size of the full data, 
and any computational benefits from subsampling wane, do the stochastic 
trajectories provide a reasonable emulation of the true trajectories.}
\label{fig:level_sets}
\end{figure}

When the full data are used, numerical trajectories generated by the second-order 
symplectic integrator constructed above closely follow the true trajectories 
(Figure \ref{fig:level_sets}a).  Approximating the potential with a subsample of the data 
introduces the aforementioned bias, which shifts the stochastic trajectory away from the 
exact trajectory despite negligible error from the symplectic integrator itself (Figure \ref{fig:level_sets}b).
Only when the size of each subsample approaches the full data set, and the computational
benefit of subsampling fades, does the stochastic trajectory provide a reasonable
approximation to the exact trajectory (Figure \ref{fig:level_sets}c)

As noted above, geometric considerations suggest that this bias should grow with the 
dimensionality of the target distribution.  To see this, consider running Hamiltonian
Monte Carlo, implemented with the same second-order symplectic integrator using
a step size, $\epsilon$, and a random integration time for each trajectory, 
$\tau \sim U \! \left(0, 2 \pi \right)$, on the multivariate generalization of \eqref{posterior},
\begin{equation} \label{multivariate_posterior}
\prod_{d = 1}^{D} p \! \left( \mu_{d} | \vec{x}_{d} \right),
\end{equation}
where the true $\mu_{d}$ are sampled from $\mu_{d} \sim \mathcal{N} \! \left( 0, 1 \right)$.
As a surrogate for the accuracy of the resulting samples I will use the average
Metropolis acceptance probability using the full data.

\begin{figure}
\centering
\includegraphics[width=3.25in]{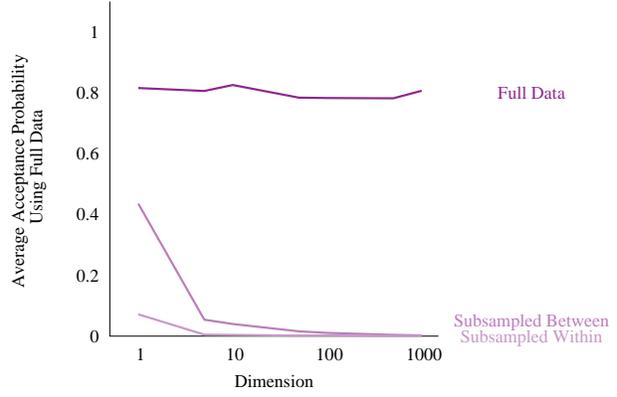}
\caption{When the full data are used, high accuracy of Hamiltonian Monte Carlo samples, 
here represented by the average Metropolis acceptance probability using the full data, can 
be maintained even as the dimensionally of the target distribution grows.  The biases induced 
when the data are subsampled, however, cannot be controlled and quickly devastate the 
accuracy of the algorithm.  Here the step size of the subsampled algorithms has been decreased
relative to the full data algorithm in order to equalize the computational cost -- even in this
simple example, a proper implementation of Hamiltonian Monte Carlo can achieve a given
accuracy much more efficiently than subsampling.}
\label{fig:multivariate}
\end{figure}

When the full data are used in this model, the step size of the symplectic integrator can be 
tuned to maintain constant accuracy as the dimensionality of the target distribution, $D$, increases.
The bias induced by subsampling between trajectories, however, is invariant to the
step size of the integrator and rapidly increases with the dimension of the target distribution.
Here the data were partitioned into $J = 25$ batches of $B = 20$ data, the subsample
used for each trajectories randomly selected from the first five batches, and the step size 
of the subsampled trajectories reduced by $N / (J \cdot B) = 5$ to equalize the computational
cost with full data trajectories (Figure \ref{fig:multivariate}).

\subsection{Subsampling Data within a Single Trajectory}

Given that using a single subsample for an entire trajectory introduces an
irreducible bias, we might next consider subsampling at each step within
a single trajectory, hoping that the bias from each subsample cancels
in expectation.  Ignoring any Metropolis correction, this is exactly the
\textit{naive stochastic gradient Hamiltonian Monte Carlo} of \cite{ChenEtAl:2014}.

To understand the accuracy of this strategy consider building up such a 
stochastic trajectory one step at a time.  Given the first two randomly-selected 
subsamples, $V_{i}$ and then $V_{j}$, the first two steps of the resulting
integrator are given by
\begin{align*}
\phi^{H_{j}}_{\epsilon} \circ \phi^{H_{i}}_{\epsilon}
&=
e^{ \epsilon \hat{H} - \epsilon  \widehat{ \Delta V}_{j} } 
\circ e^{ \epsilon \hat{H} - \epsilon  \widehat{ \Delta V}_{i} }
+ \mathcal{O} \! \left( \epsilon^{3} \right)
\\
&=
\exp \! \left( 
2 \epsilon \hat{H} - \epsilon \left(  \widehat{ \Delta V}_{i} +  \widehat{ \Delta V}_{j} \right)
\right.
\\
& \hspace{10mm} \left.
+ \frac{\epsilon^{2}}{2} \left[ \hat{H} -  \widehat{ \Delta V}_{j}, \hat{H} - \widehat{ \Delta V}_{i} \right]
\right)
+ \mathcal{O} \! \left( \epsilon^{3} \right)
\\
&=
\exp \! \left( 
2 \epsilon \hat{H} - \epsilon \left( \widehat{ \Delta V}_{i} + \widehat{ \Delta V}_{j} \right)
\right.
\\
& \hspace{10mm} \left.
+ \frac{\epsilon^{2}}{2} \left(
- \left[ \hat{H}, \hat{V}_{\setminus i} \right]
- \left[ \hat{V}_{\setminus j}, \hat{H} \right] \right) \right)
+ \mathcal{O} \! \left( \epsilon^{3} \right),
\end{align*}
where we have used the fact that the $\{ \widehat{ \Delta V}_{j} \}$ commute
with each other.  Similarly, the first three steps are given by
\begin{align*}
\phi^{H_{k}}_{\epsilon} \circ \phi^{H_{j}}_{\epsilon} \circ \phi^{H_{i}}_{\epsilon}
&
\\
& \hspace{-15mm} =
\exp \! \left( 
3 \epsilon \hat{H} 
- \epsilon \left( \widehat{ \Delta V}_{i} +  \widehat{ \Delta V}_{j} +  \widehat{ \Delta V}_{k} \right)
\right.
\\
& \hspace{-8mm} \left.
+ \frac{\epsilon^{2}}{2} \left(
- \left[ \hat{H}, \widehat{ \Delta V}_{i} \right]
- \left[ \widehat{ \Delta V}_{j}, \hat{H} \right] \right) \right.
\\
& \hspace{-8mm} \left.
+ \frac{\epsilon^{2}}{2} \left(
\left[ \hat{H} - \widehat{ \Delta V}_{k}, 2 \hat{H} - \widehat{ \Delta V}_{i} - \widehat{ \Delta V}_{j} \right]
\right)
\right)
\\
& \hspace{-11mm} + \mathcal{O} \! \left( \epsilon^{3} \right)
\end{align*}
%
\begin{align*}
\phi^{H_{k}}_{\epsilon} \circ \phi^{H_{j}}_{\epsilon} \circ \phi^{H_{i}}_{\epsilon}
\\
& \hspace{-15mm} =
\exp \! \left( 
3 \epsilon \hat{H} 
- \epsilon \left( \widehat{ \Delta V}_{i} + \widehat{ \Delta V}_{j} + \widehat{ \Delta V}_{k} \right)
\right.
\\
& \hspace{-8mm} \left.
- \epsilon^{2} \left( \left[ \hat{H}, \widehat{ \Delta V}_{i} \right] 
- \left[ \hat{H}, \widehat{ \Delta V}_{k} \right] \right)
\right) + \mathcal{O} \! \left( \epsilon^{3} \right),
\end{align*}
and, letting $j_{l}$ denote the subsample chosen at the $l$-th step, the composition over
an entire trajectory becomes
\begin{align*}
\circ_{l = 1}^{L} \phi^{H_{j_{l}}}_{\epsilon}
& \\
& \hspace{-10mm} =
\exp \! \left( 
\left( L \epsilon \right) \hat{H} 
- \left( L \epsilon \right) \frac{1}{L} \sum_{l = 1}^{L}  \widehat{ \Delta V}_{j_{l}}
\right.
\\
& \hspace{2mm} \left.
+ \left( L \epsilon \right) \epsilon 
\left( \left[ \hat{H}, \widehat{ \Delta V}_{j_{1}} \right] 
- \left[ \hat{H}, \widehat{ \Delta V}_{j_{l}} \right] \right) 
\right)
\\
& \hspace{-7mm}
+ \left(L \epsilon \right) \mathcal{O} \! \left( \epsilon^{2} \right)
\\
& \hspace{-10mm} =
\exp \! \left( \tau \hat{H} - \tau \frac{1}{L} \sum_{l = 1}^{L} \widehat{ \Delta V}_{j_{l}}
\right.
\\
& \hspace{2mm} \left.
+ \tau \epsilon 
\left( \left[ \hat{H}, \widehat{ \Delta V}_{j_{1}} \right] 
- \left[ \hat{H}, \widehat{ \Delta V}_{j_{L}} \right] \right)
\right)
+ \mathcal{O} \! \left( \epsilon^{2} \right)
\\
& \hspace{-10mm} =
\exp \! \left( \tau \hat{H} + \tau B_{1} + \tau B_{2} \right)
+ \mathcal{O} \! \left( \epsilon^{2} \right),
\end{align*}
where
\begin{equation*}
B_{1} = - \frac{1}{L} \sum_{l = 1}^{L} \widehat{ \Delta V}_{j_{l}}
\end{equation*}
and
\begin{equation*}
B_{2} = \epsilon 
\left( \left[ \hat{H}, \widehat{ \Delta V}_{j_{1}} \right] 
- \left[ \hat{H}, \widehat{ \Delta V}_{j_{L}} \right] \right).
\end{equation*}
Once again, subsampling the data introduces bias into the numerical trajectories.

Although the second source of bias, $B_{2}$, is immediately rectified by appending the 
stochastic trajectory with an update from the initial subsample such that $j_{L} = j_{1}$, 
the first source of bias, $B_{1}$, is not so easily remedied.  Expanding,
\begin{align*}
\frac{1}{L} \sum_{l = 1}^{L}  \widehat{ \Delta V}_{j_{l}}
&=
\frac{1}{L} \sum_{l = 1}^{L} \left( \hat{V} - J \, \hat{V}_{j_{l}} \right)
\\
&=
\hat{V} - \frac{J}{L} \sum_{n = 1}^{L} \hat{V}_{j_{l}}
\\
&=
- \left( \frac{ \partial V }{ \partial q }  - \frac{J}{L} \sum_{l = 1}^{L} \frac{ \partial V_{j} }{ \partial q }
\right) \frac{ \partial }{ \partial p },
\end{align*}
we see that $B_{1}$ vanishes only when the average gradient of the selected 
subsamples yields the gradient of the full potential.  Averaging over subsamples 
may reduce the bias compared to using a single subsample over the entire
trajectory \eqref{eqn:between_bias}, but the bias still scales poorly with the 
dimensionality of the target distribution (Figure \ref{fig:subsampled_gradients}).

In order to ensure that the bias vanishes identically and independent of the redundancy
of the data, we have to use each subsample the same number of times within a single 
trajectory.  In particular, both biases vanish if we use each subsample twice in a symmetric 
composition of the form
\begin{equation*}
\left( \circ_{l = 1}^{L} \phi^{H_{l}}_{\epsilon} \right)
\circ 
\left( \circ_{l = 1}^{L} \phi^{H_{L + 1 - l}}_{\epsilon} \right).
\end{equation*}
Because this composition requires using all of the subsamples it does not provide
any computational savings and it seems rather at odd with the original stochastic 
subsampling motivation. 

\begin{figure}
\centering
\subfigure[]{\includegraphics[width=1.95in]{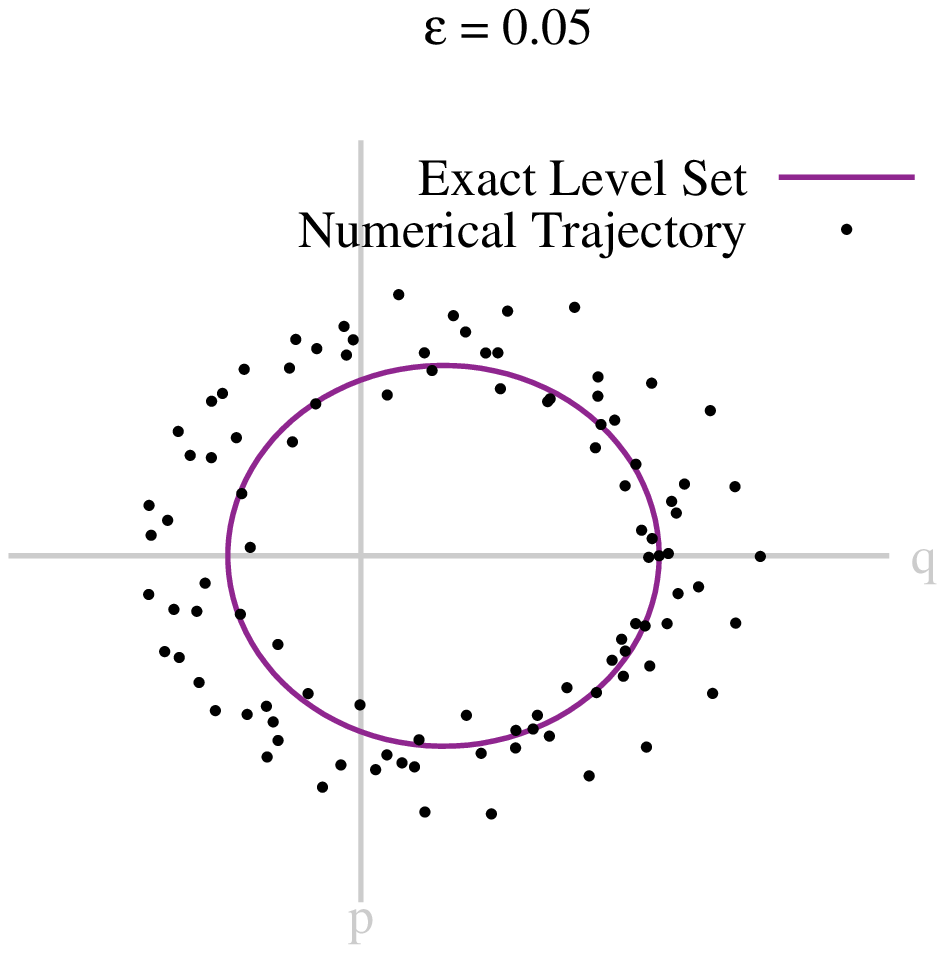}}
\subfigure[]{\includegraphics[width=1.95in]{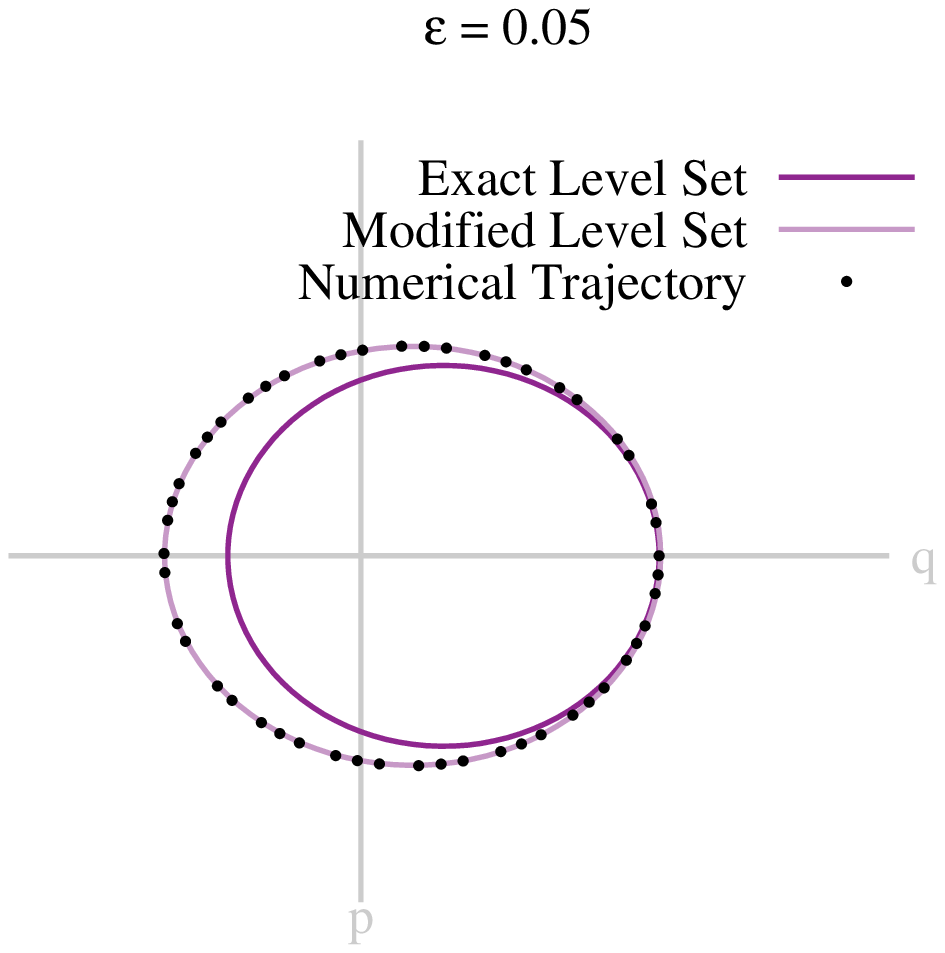}}
\subfigure[]{\includegraphics[width=1.95in]{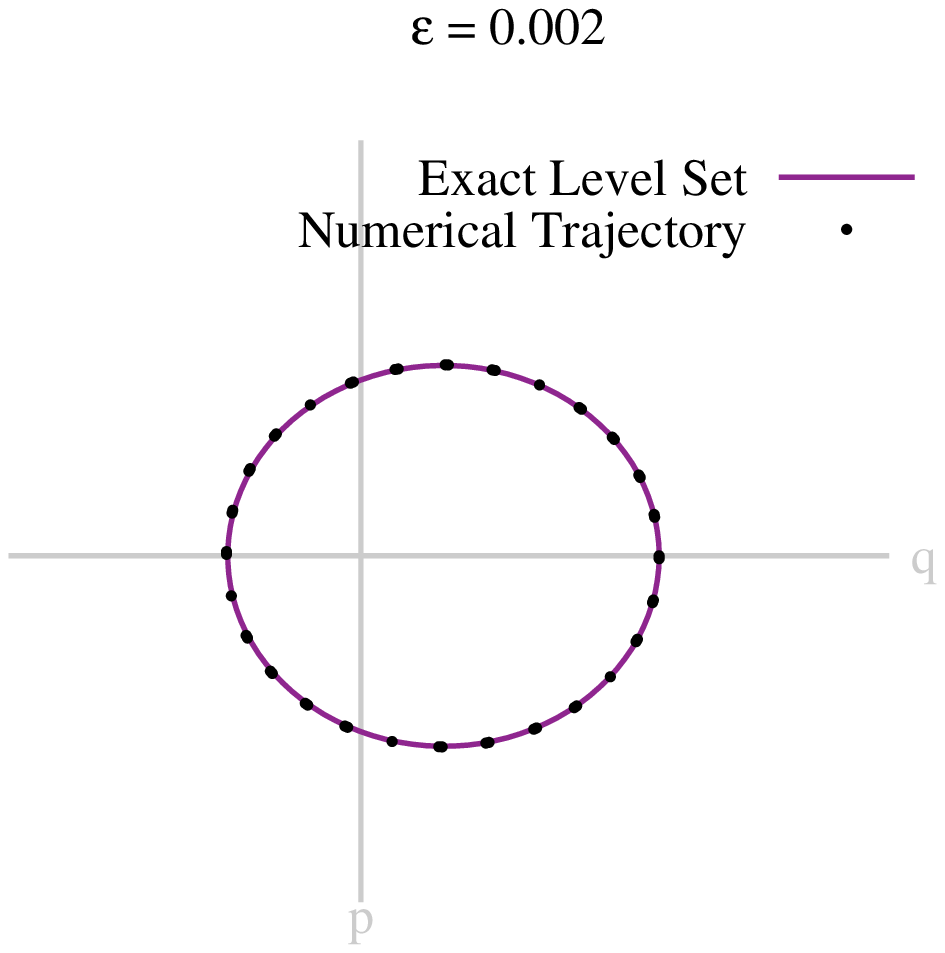}}
\caption{The symmetric composition of flows from each subsamples of the data 
eliminates all bias in the stochastic trajectory because it implicitly reconstructs a 
symplectic integrator.  Refining (a) all intermediate steps in a stochastic trajectory 
(b) to only those occurring after a symmetric sweep of the subsamples reveals
the level set of the modified Hamiltonian corresponding to the implicit
symplectic integrator.  Because of the vanishing bias, (c) the error in the
stochastic trajectory can be controlled by taking the step size to zero. }
\label{fig:symmetric_stochastic}
\end{figure}

Indeed, this symmetric composition is not stochastic at all and actually corresponds 
to a rather elaborate symplectic integrator with an effective step size of $J \epsilon$, 
where the potential energy from each subsample generates its own flow, equivalent
to the integrator in Split Hamiltonian Monte Carlo \cite{ShahbabaEtAl:2014}.  Removing 
intermediate steps from this symmetric, stochastic trajectory 
(Figure \ref{fig:symmetric_stochastic}a) reveals the level set of the corresponding
modified Hamiltonian (Figure \ref{fig:symmetric_stochastic}b).  Because this
symmetric composition integrates the full Hamiltonian system, the error is 
once again controllable and vanishes as the step size is decreased
(Figure \ref{fig:symmetric_stochastic}c).

\begin{figure}
\centering
\subfigure[]{\includegraphics[width=2.9in]{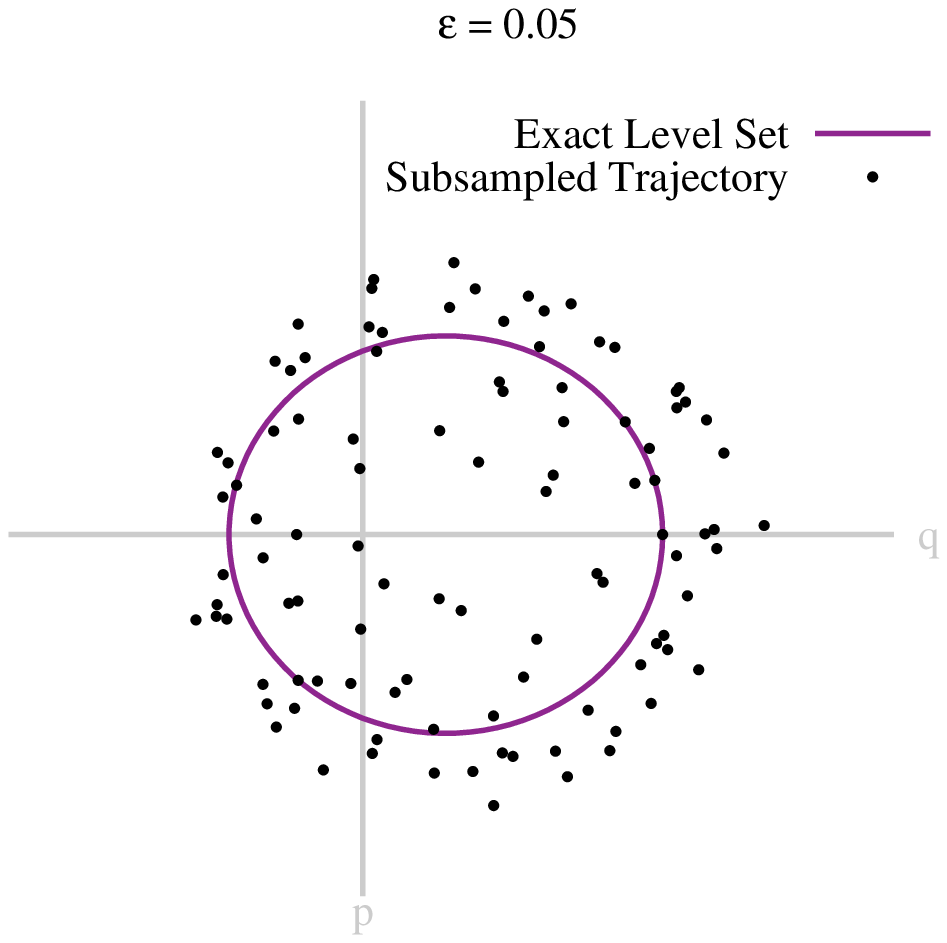}}
\subfigure[]{\includegraphics[width=2.9in]{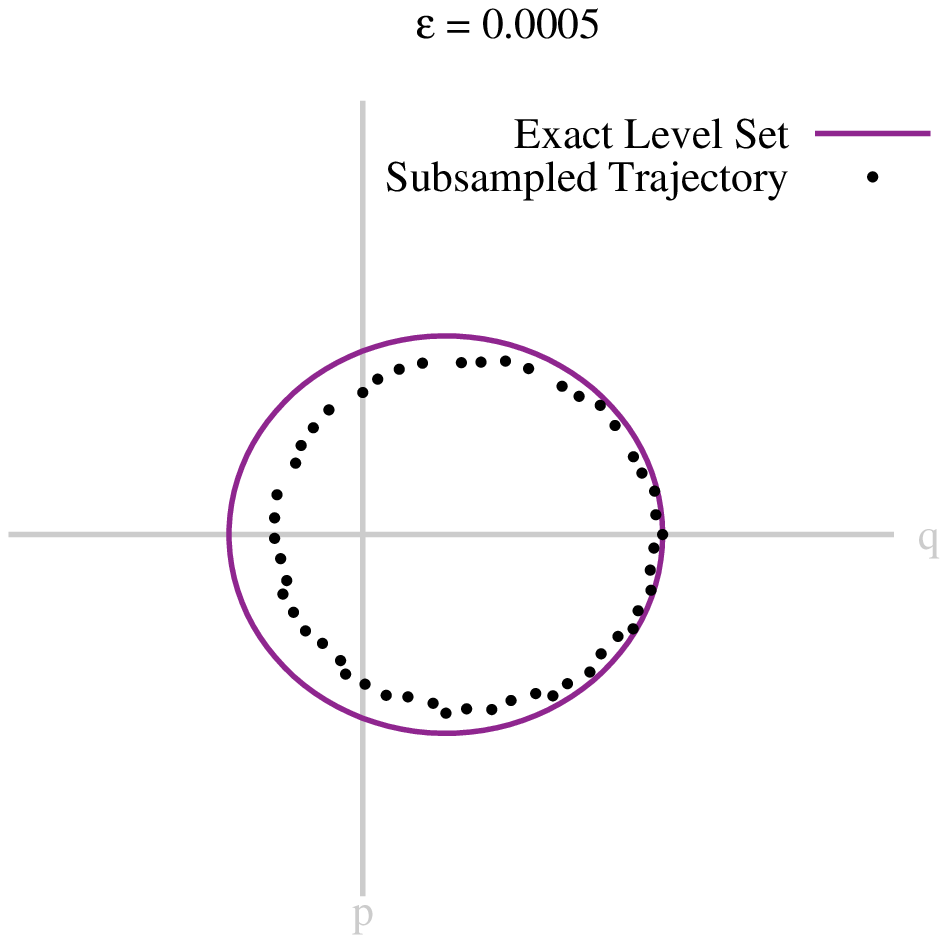}}
\caption{(a) Utilizing only a few subsamples within a trajectory yields numerical 
trajectories biased away from the exact trajectories.  (b) Unlike the error introduced 
by a full symplectic integrator, this bias is irreducible and cannot be controlled
by tuning the step size.  The performance of such an algorithm is limited by 
the size of the bias which itself depends on the redundancy of the data relative
to the target model.}
\label{fig:subsample_trajectory}
\end{figure}

Limiting the number of subsamples, however, leaves the irreducible bias
in the trajectories that cannot be controlled by the tuning the step size
(Figures \ref{fig:multivariate}, \ref{fig:subsample_trajectory}).  Once more 
we are left dependent on the redundancy of the data for any hope of improved 
performance with subsampling.

\section{Conclusion}

The efficacy of Markov Chain Monte Carlo for complex, high-dimensional 
target distributions depends on the ability of the sampler to explore the intricate 
and often meandering neighborhoods on which probability is distributed.
Symplectic integrators admit a structure-preserving implementation of
Hamiltonian Monte Carlo that is amazingly robust to this complexity
and capable of efficiently exploring the most complex target distributions.
Subsampled data, however, does not in general have enough information 
to enable such efficient exploration; this lack of information manifests as
an irreducible bias that devastates the scalable performance of Hamiltonian 
Monte Carlo.

Consequently, without having access to the full data there is no immediate way of 
engineering a well-behaved implementation of Hamiltonian Monte Carlo applicable
to most statistical models.  As with so many other subsampling algorithms, the 
adequacy of a subsampled Hamiltonian Monte Carlo implementation is at the mercy 
of the redundancy of the data relative to the complexity of the target model, and not 
in the control of the user.

Unfortunately many of the problems at the frontiers of applied statistics are 
in the \textit{wide data} regime, where data are sparse relative to model 
complexity.  Here subsampling methods have little hope of success;
we must focus our efforts not on modifying Hamiltonian Monte Carlo but rather
on improving its implementation with, for example, better memory management 
and efficiently parallelized gradient calculations.



\bibliography{stochastic_integrators}
\bibliographystyle{icml2015}

\end{document}